\documentclass[dvips]{article}

\usepackage{icrc2011}
\usepackage{amssymb}
\usepackage{marvosym}

\def\snr{SNR\,G330.2$+$1.0}
\def\name{G330.2$+$1.0}

\def\snrthree{SNR\,G330.2$+$1.0}
\def\gthree{G330.2$+$1.0}
\def\snrone{SNR\,G1.9$+$0.3}
\def\gone{G1.9$+$0.3}

\def\gammaray{$\gamma$-ray}
\def\gammarays{$\gamma$-rays}
\def\hess{H.E.S.S.}

\renewcommand{\deg}{\mbox{\ensuremath{^\circ}}} 
\newcommand{\arcmin}{\mbox{\ensuremath{^{\prime}}}}

\title{VHE gamma-ray observations of the young synchrotron-dominated SNRs G1.9+0.3 and G330.2+1.0 with H.E.S.S.}

\newcommand{\etal}{\MakeLowercase{\textit{et al. }}} 
\shorttitle{Sushch \etal H.E.S.S. observations of SNRs G1.9+0.3 and G330.2+1.0}

\authors{Iurii Sushch$^{1,2\,\textrm{\Email}}$, Ryan C. G. Chaves$^{3}$, Manuel Paz Arribas$^{1,4}$, Francesca Volpe$^{3}$, Nukri Komin$^{5}$, Matthias Kerschhaggl$^{6}$ for the H.E.S.S. Collaboration}

\afiliations{$^1$Humboldt University of Berlin, Germany\\ $^2$Taras Shevchenko National University of Kyiv, Ukraine
\\$^3$Max Planck Institute for Nuclear Physics, Heidelberg, Germany\\$^4$DESY, D-15735 Zeuthen, Germany\\$^5$LAPP, CNRS/IN2P3, Annecy-le-Vieux, France
\\$^6$University of Bonn, Germany}
\email{\Email \,yusushch@physik.hu-berlin.de}

\abstract{Supernova remnants (SNRs) are widely considered to be accelerators of 
cosmic rays (CR). They are also expected to produce very-high-energy
(VHE; $E > 100$ GeV) gamma rays through interactions of high-energy
particles with the surrounding medium and photon fields. They are,
therefore, promising targets for observations with ground-based
imaging atmospheric Cherenkov telescopes like the H.E.S.S. telescope
array. VHE gamma-ray emission has already been discovered from a
number of SNRs, establishing them as a prominent source class in the
VHE domain. Of particular interest are the handful of SNRs whose X-ray
spectra are dominated by non-thermal synchrotron emission, such as the
VHE gamma-ray emitters RX J0852.0-4622 (Vela Jr.) and RX J1713-3946.
The shell-type SNRs G1.9+0.3 and G330.2+1.0 also belong to this
subclass and are further notable for their young ages ($\leq 1$ kyr),
especially G1.9+0.3, which was recently determined to be the youngest
SNR in the Galaxy ($\sim100$ yr). These unique characteristics motivated
investigations with H.E.S.S. to search for VHE gamma rays. The results
of the H.E.S.S. observations and analyses are presented, along
with implications for potential particle acceleration scenarios.}

\begin{document}
\maketitle

\section{Introduction}
Supernova remnants (SNRs) are believed to be sites of efficient 
particle acceleration and are
expected to produce very-high-energy 
(VHE; $E > 0.1$\,TeV) \gammarays\ through the interaction of these accelerated, high-energy particles with 
ambient media and fields. VHE \gammaray\ emission is currently detected from a number of SNRs. 
Of particular interest are those 
SNRs whose X-ray spectra are dominated by non-thermal emission 
such as RX\,J0852.0$-$4622 (Vela Jr.) \cite{Aharonian07VelaJr} and 
RX\,J1713$-$3946 \cite{Aharonian07RXJ1713}. 
Synchrotron emission from these SNRs reveals the existence of high-energy 
electrons implying strong particle acceleration at 
shock fronts of remnants. High-energy particles accelerated at shock fronts can produce 
VHE \gammarays\ through the inverse Compton (IC) scattering of relativistic electrons on 
ambient photon fields and/or through proton-proton interactions.

In this paper the results of the H.E.S.S. observations of two other SNRs with predominantly non-thermal 
X-ray emission, \snrone\ \cite{reynolds08} and \snr\ \cite{torii06}, are 
presented.

\section{SNRs \gone\ and \gthree}

\subsection{\gone}
\gone\ is the youngest known Galactic supernova remnant. An age of about 
100 years was derived from the expansion rate of the object obtained from the comparison
of radio observations in 1985 and Chandra observations in 2007 \cite{reynolds08}. 
This result was confirmed by independent radio observations \cite{green08, murphy08}.

\gone\ was first identified as an SNR in a radio survey using the Very Large Array (VLA) 
at 4.9\,GHz based on its shell-like morphology and non-thermal radio emission \cite{green84}. 
\gone\ had the smallest angular extent ever measured for a Galactic SNR ($\sim$1.2\arcmin). 

Chandra observations 22 years later showed a significant expansion of the remnant to 
$1.7^{\prime}$ in diameter (about 16\%)
\cite{reynolds08}. The X-ray image shows a nearly circular bright ring with 
extensions (``ears'') extruding symmetrically from the east and west sides. There is a significant difference 
in radio and X-ray morphologies: while the radio source has a brightness maximum to the north, 
the X-ray image reveals a bilateral east-west symmetry.   
Observations revealed a featureless synchrotron-dominated X-ray spectrum 
\cite{reynolds08} indicating the presence of a high-energy electron population. 
In the framework of the \texttt{srcut} model, which describes the synchrotron 
radiation from the exponentially cut-off power-law electron distribution, the roll-off frequency $\nu_{\mathrm{roll}} 
= 1.4 \times 10^{18}$ Hz (one of the highest values ever reported for SNRs) and the radio spectral 
index $\alpha = 0.65$ were obtained. The very high column density of $N_{\rm{H}} = 
5.5\times10^{22}$ cm$^{-2}$ suggests a distance of about 8.5 kpc. At this distance, the observed 
angular radius of the bright X-ray ring corresponds to $\sim 2$ pc, or 2.2 pc including  the 
``ears''. The observed expansion of the remnant leads to an estimate of the 
shock speed of 14000 km/s. For a sphere of radius 2.2 pc an explosion model with an exponential 
ejecta profile \cite{dwarkadas&chevalier} predicts an age of 100 years and an ISM number 
density of about 0.04 cm$^{-3}$. Slightly different values of the 
age, 80 years, and a number density of 0.018 cm$^{-3}$ are derived in \cite{ksenofontov10}, assuming an expansion velocity of 14000 km/s 
and a radius of 2 pc.

\subsection{\gthree}
The radio source \name\ was identified as a Galactic SNR in \cite{clark73} and 
\cite{clark75} on the basis of its non-thermal spectrum and its proximity to 
the galactic plane. Subsequent observations at radio frequencies \cite{caswell83} 
showed the shell-like structure of the remnant with a gradient of surface brightness towards the galactic plane. 
\gthree\ was classified as a possible composite type SNR in \cite{whiteoak&green96}. The angular diameter of the shell 
is $\sim11^{\prime}$ \cite{caswell83, whiteoak&green96}.  

Based on ASCA observations \cite{tanaka94}, a featureless
X-ray spectrum with a photon index of $\Gamma \simeq 2.8$ and interstellar column density 
$N_{\rm{H}}\simeq2.6\times10^{22}$ cm$^{-2}$ was discovered \cite{torii06}. The X-ray spectrum was also fit with the \texttt{srcut} model deriving a roll-off frequency $\nu_{\mathrm{roll}} = 
4.3 \times 10^{15}$ Hz and an absorption column density $N_{\rm{H}}\simeq5.1\times10^{22}$ 
cm$^{-2}$ for the assumed radio spectral index $\alpha=0.3$. X-ray observations revealed a general anticorrelation 
between radio and X-ray intensities. 

Subsequent Chandra and XMM-Newton observations \cite{park06, park09} revealed that 
the X-ray emission from \snr\ is dominated by a power-law continuum ($\Gamma\sim2.1-2.5$) 
and comes primarily from thin filaments along the boundary shell. 
A point-like source CXOU J160103.1-513353 was discovered with Chandra \cite{park06} at the center of the SNR, 
a possible central compact object (CCO). 
Chandra and XMM-Newton observations also revealed a 
faint thermal X-ray emission from \name\ \cite{park09}. Using the thermal emission, a low ISM 
density of $0.1$ cm$^{-3}$ was calculated. A lower limit on the 
distance $d_{\rm{G330}}\geq4.9$ kpc was calculated in \cite{mcclure01} using the HI 
absorption measurement. Hereafter, the distance to \name\ is assumed to be 5 kpc. 
Obtained values of the ISM density and distance lead to the 
age estimate for the remnant of $t_{\rm{G330}}\simeq1000$ years, using the 
Sedov solution \cite{sedov} to describe the hydro-dynamical expansion of the SNR at the adiabatic stage \cite{park09}.


\section{Observations and Analysis}
\subsection{The H.E.S.S. Instrument}

H.E.S.S. (High Energy Stereoscopic System) is an array of four 13 m diameter imaging atmospheric
Cherenkov telescopes (IACTs) located in the Khomas Highland of Namibia
at an altitude of 1800 m above sea level \cite{bernloehr03,
funk04}.The telescopes are optimized for detection of very high energy $\gamma$-rays in the 
range of ~100 GeV to 20 TeV by imaging Cherenkov light emitted by charged particles in an
Extensive Air Shower (EAS). The total field of view of H.E.S.S. is $5^{\circ}$ in diameter. The angular 
resolution of the system is $\lesssim 0.1^{\circ}$ and the average energy resolution is about 
$15\%$ \cite{crab}. The H.E.S.S. array is capable to detect point sources with a flux of $1\%$ 
of the Crab Nebula flux at the significance level of $5\sigma$ in 25 hours at low zenith angles \cite{crab}. 

\subsection{Data Set and Analysis Results}

\begin{table*}[ht]
\centering
\caption{\hess\ observations of SNRs \gone\ and \gthree}
\label{data}
\begin{tabular}{c c c c c c}
\\
\hline
\hline
\\
SNR & Observation period & Livetime & Mean offset angle & Mean zenith angle & Threshold energy \\
\hline
\\
\gone\   & 2004--2010 & 75\,h & $1.5\deg$ & $19\deg$ & 0.26\,TeV \\
\gthree\ & 2004--2010                          & 20\,h        & 1.3\deg               & 32\deg              & 0.38\,TeV \\
%
%
%
%
\hline
\end{tabular}
\end{table*}

For the study of \gone\ and \name, data taken in the period 
from 2004 to 2010 were compiled. \snrone\ is located $\sim2$\deg from the supermassive black hole 
Sgr A$^{*}$ at the Galactic Center (GC). More than a half of the observations used for the 
analysis are therefore obtained from Sgr A$^{*}$ observations. In order to reduce the high exposure gradient 
towards the GC, only those observations with a pointing within 2.0\deg\ from \gone\ were selected for 
the analysis. For \snrthree\ all pointings within 2.5\deg\ from the center of the remnant were considered. 
After the standard H.E.S.S. quality selection 
\cite{crab} this resulted in 75 and 20 hours of livetime for \gone\ and \name, 
respectively (see Table~\ref{data}). 

The H.E.S.S. standard Hillas reconstruction \cite{aharonian06} with standard cuts 
and the Reflected Region Background method \cite{berge07} were used for the data analysis.
The results were cross-checked using the alternative \texttt{Model analysis} 
technique \cite{deNaurois09} as well as independent 
calibration of the raw data and quality selection criteria.

\begin{figure}[h]
\centering
\resizebox{\hsize}{!}{\includegraphics{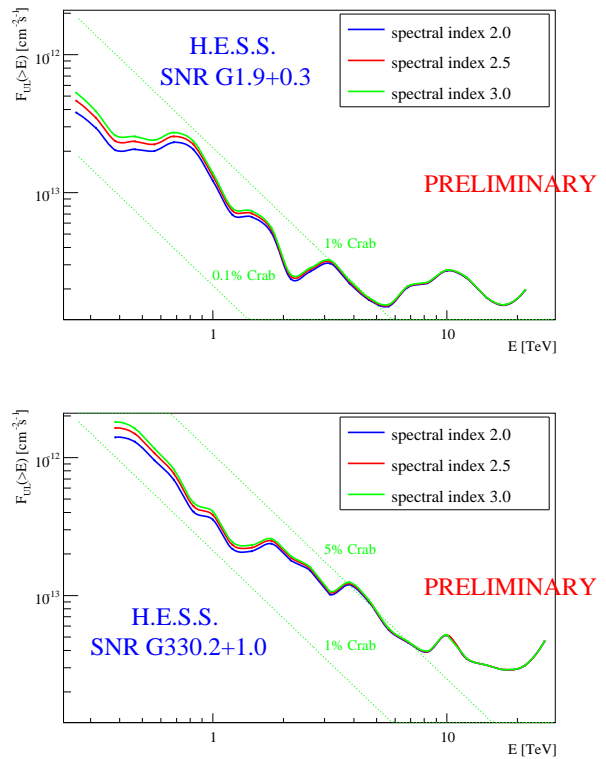}}
\caption{The upper limit (99\% confidence level) of the integrated TeV \gammaray\ flux from SNRs \gone\ (top) and \gthree\ (bottom) 
for three different assumed spectral indices, $\Gamma = 2.0,$ $2.5\mathrm{ and }3.0$.}
\label{UL_G1.9_and_G330}
\end{figure}

No significant $\gamma$-ray signal was detected from \gone\ or \name. 
The $99 \%$ confidence level upper limits \cite{upper_limit} 
on the integrated fluxes above 260 GeV (\gone) and 380 GeV (\name) energy thresholds were calculated 
for three different assumed spectral indices, 2.0, 2.5 and 3.0 (see Table~\ref{UL}).
The assumed spectral index does not have a major impact in the upper limits (see Fig.~\ref{UL_G1.9_and_G330}).

\begin{table}
\tabcolsep 5.8pt
\small
\caption{Upper limits on the TeV \gammaray\ flux from SNRs \gone\ and \gthree}
\label{UL}
\centering
\begin{tabular}{@{}c c c@{}}
\\
\hline
\hline
&& \\
Spectral&  $F_{\rm{G1.9+0.3}}(E>260\hbox{ GeV})$,  &   $F_{\rm{G330+1.0}}(E>380\hbox{ GeV})$, \\
index& [cm$^{-2}$ s$^{-1}$] & [cm$^{-2}$ s$^{-1}$] \\
\hline
&&\\
2.0           & $<3.8\times10^{-13}$     & $<1.4\times10^{-12}$\\

2.5           & $<4.6\times10^{-13}$     & $<1.6\times10^{-12}$\\

3.0           &  $<5.3\times10^{-13}$    & $<1.8\times10^{-12}$\\


\hline

\end{tabular}
\end{table}

\section{Discussion}
  VHE $\gamma$-rays from SNRs can be produced either via inverse Compton (IC)
  scattering of relativistic electrons on photon fields or via proton-proton interactions. In this 
  paper the first scenario is discussed.
 
  The radio and X-ray data were fit with the model of the synchrotron emission of an electron population 
  with an energy distribution which follows a power-law with an exponential cut-off
\begin{equation}
\frac{dN_\mathrm{e}}{dE}\propto E^{-p}e^{-E/E_\mathrm{cut}}. 
\end{equation}
  The distribution of relativistic electrons can be completely described by the set of 
  three parameters: the spectral index $p$, the cut-off energy $E_\mathrm{cut}$ and the total 
  energy in electrons $W_\mathrm{tot}$. These parameters can be determined from the fit of 
  the radio and X-ray data if the magnetic field $B$ is known. Independently of the magnetic field, 
  the fit of the synchrotron emission spectrum to radio and X-ray data can be defined by the similar set of three 
  parameters: spectral index $\alpha$, roll-off energy $E_\mathrm{roll}$ and the spectrum normalization 
  at 1 GHz. The electron spectral index $p$ is directly connected to $\alpha$ through the expression $p = 2\alpha + 1$, 
  while $E_\mathrm{cut}$ and $W_\mathrm{tot}$ together with the synchrotron spectrum parameters depend also on 
  the magnetic field.

\begin{figure}[h]
    \centering
    \resizebox{\hsize}{!}{\includegraphics{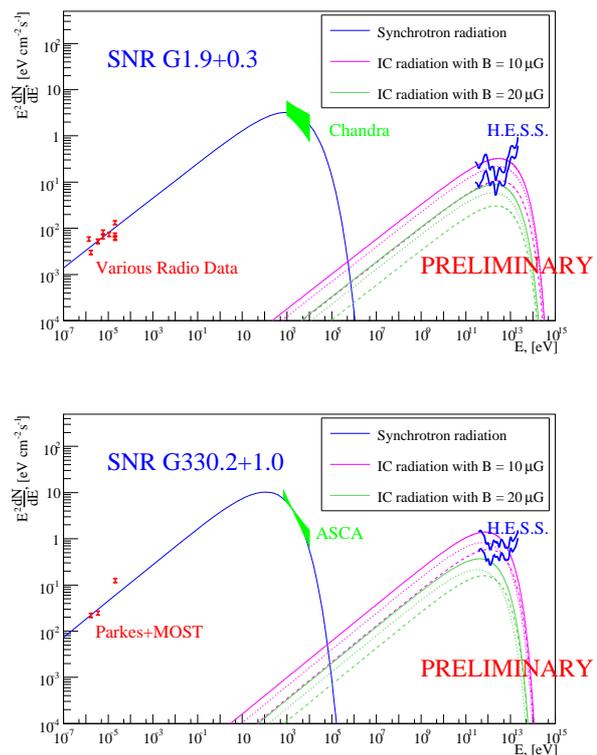}}
    \caption{Spectral energy distribution (SED) of the broadband emission from the SNRs \gone\ and \gthree. The H.E.S.S. upper limits are shown assuming two different spectral indices: 2.0 (lower curve) and 3.0 (upper curve). The various radio data for \gone\ were compiled in \cite{green08}. For the IC emission: dashed lines correspond to the IC on IR photons, dotted lines to the IC on CMB photons, solid lines to the total IC emission}
    \label{SED}
  \end{figure}

  Assuming different values for the magnetic field, the gamma-ray emission created by the same 
  population of electrons via IC scattering on ambient photon fields 
  (infrared (IR) and CMB) is predicted (Fig~\ref{SED}). The photon distribution is derived from the 
  interstellar radiation field (ISRF) model \cite{porter06} and is fit with 
  three Planck distributions of optical, IR and CMB photons. The contribution of 
  the optical photons to the \gammaray\ flux is found to be negligible.
  The upper limit on the TeV flux 
  leads to an estimate of a lower limit of the magnetic field. This in turn 
  allows to determine upper limits on the electron cut-off energy and the total energy in 
  electrons (see Table~\ref{param}). 

  Recently, the first synchrotron-dominated 
  SNR RX J1713.7-3946 was discovered in GeV $\gamma$-rays \cite{1713}. The observed spectrum 
  favours the leptonic model and implies a very low magnetic field of $\simeq$10 $\mu$G which is 
  comparable to the limits found for \gone\ and \name. 
 
  \begin{table}
    \centering
    \caption{SED model fitting parameters}
    \label{param}
    \begin{tabular}{ccc}
      \\
      \hline
      \hline
      SNR&\gone\ & \name\ \\
      \hline
      $p$& 2.2 & 2.2\\
      $E_\mathrm{roll}$, [KeV]& 2.0& 0.2\\
      $B$, [$\mu$G]&$>15$&$>14$\\
      $E_\mathrm{cut}$, [TeV]& $<50$& $<17$\\
      $W_\mathrm{tot}$, [erg] & $<3.1\times10^{48}$& $<5.7\times10^{48}$\\ 
      \hline
    \end{tabular}
  \end{table}

\section{Summary}
Results from H.E.S.S. observations of two SNRs with dominantly non-thermal X-ray spectra, \gone\ and \gthree, were presented. 
$99 \%$ confidence level upper limits on the TeV flux from these sources were estimated. 
For an assumed spectral index of 2.5, the upper limits are found to be 
$F_{\rm{G\,1.9}}(>260\hbox{ GeV})<4.6\times10^{-13}$ cm$^{-2}$s$^{-1}$ for \gone\ 
and $F_{\rm{G\,330}}(>380\hbox{ GeV})<1.6\times10^{-12}$ cm$^{-2}$s$^{-1}$ for \gthree. 

Upper limits on the VHE flux allow to derive constraints on the interior magnetic field in the 
framework of a leptonic emission scenario. The obtained lower limits on the magnetic fields are 
comparable to the estimates of magnetic fields 
in SNRs in which VHE emission can be explained by an IC mechanism.

\gthree\ and \gone\ remain promising targets for observations at VHE and could be detectable 
with future, more sensitive instruments like the Cherenkov Telescope Array (CTA).


\begin{thebibliography}{}
\bibitem{Aharonian07VelaJr} F. Aharonian et al., ApJ, 2007, {\bf661}: 236-249 
\bibitem{Aharonian07RXJ1713} F. Aharonian et al.,  A\&A, 2007, {\bf464}: 235-243
\bibitem{reynolds08} S. P. Reynolds et al., ApJ, 2008, {\bf680}: L41--L44
\bibitem{torii06} K. Torii et al., PASJ, 2006, {\bf58}: L11--L14
\bibitem{green08} D. A. Green et al., MNRAS, 2008, {\bf387}: L54--L58
\bibitem{murphy08} T. Murphy, B. M. Gaensler, S. Chatterjee, MNRAS, 2008, {\bf389}: L23--L27
\bibitem{green84} D. A. Green, S. F. Gull, Nature, 1984, {\bf312}: 527--529
\bibitem{dwarkadas&chevalier} V. V. Dwarkadas, R. A. Chevalier, ApJ, 1998, {\bf497}: 807
\bibitem{ksenofontov10} L. T. Ksenofontov, H. J. V\"{o}lk, E. G. Berezhko, ApJ, 2010, {\bf714}: 1187--1193
\bibitem{clark73} D. H. Clark, J. L. Caswell, A. J. Green, Nature, 1973, {\bf246}: 27-30 
\bibitem{clark75} D. H. Clark, J. L. Caswell, A. J. Green, Aust. J. Phys. Astrophys. Suppl., 1975, {\bf37}: 1-38 
\bibitem{caswell83} J. L. Caswell et al., MNRAS, 1983, {\bf204}: 915--920
\bibitem{whiteoak&green96} J. B. Z. Whiteoak, A. J. Green, A\&AS, 1996, {\bf118}: 329--380
\bibitem{tanaka94} Y. Tanaka, H. Inoue, S.S. Holt, PASJ, 1994, {\bf46}: L37
\bibitem{park06} S. Park et al., ApJ, 2006, {\bf653}: L37--L40
\bibitem{park09} S. Park et al., ApJ, 2009, {\bf695}: 431--441
\bibitem{mcclure01} N. M. McClure-Griffiths et al., ApJ, 2001, {\bf551}: 394--412
\bibitem{sedov} L. I. Sedov: 1959, Similarity and Dimensional Methods in Mechanics, Academic Press, New York
\bibitem{bernloehr03} K.~Bernloehr et al., Astropart. Phys., 2003, {\bf20}: 111
\bibitem{funk04} S. Funk et al., Astropart. Phys., 2004, {\bf22}: 285
\bibitem{crab} F. Aharonian et al., A\&A, 2006, {\bf457}: 899--917
\bibitem{aharonian06} Aharonian, F., Akhperjanian, A.~G., Bazer-Bachi, A.~R., {et~al.} 2006, A\&A,
  457, 899
\bibitem{berge07} {Berge}, D., {Funk}, S., \& {Hinton}, J. 2007, A\&A, 466, 1219
\bibitem{deNaurois09} M. de Naurois, L. Rolland, Astropart. Phys., 2009, {\bf32}: 231-252
\bibitem{upper_limit} Feldman, G.~J. \& Cousins, R.~D. 1998, PHYS.REV.D, 57, 3873
\bibitem{porter06} {Porter}, T.~A., {Moskalenko}, I.~V., \& {Strong}, A.~W. 2006, ApJ, 648, L29
\bibitem{1713} Abdo, A.~A., {et~al.} 2011, ApJ, 734, 28




\end{thebibliography}



\clearpage
\end{document}